\documentclass[preprintnumbers, floatfix, twocolumn,
preprintnumbers, letterpaper, superscriptaddress,nofootinbib]{revtex4}

\usepackage{graphicx}
\usepackage{microtype}
\usepackage{amsmath}
\usepackage{amssymb}
\usepackage{subfigure}
\usepackage{hyperref}
\usepackage{url}
\usepackage{xcolor}
\usepackage{color}
\usepackage{mathrsfs}
\usepackage{amsfonts}
\usepackage{latexsym}
\usepackage{epsfig}

\definecolor{vividviolet}{rgb}{0.62, 0.0, 1.0}
\definecolor{amaranth}{rgb}{0.9, 0.17, 0.31}
\definecolor{palatinateblue}{rgb}{0.15, 0.23, 0.89}
\definecolor{brightpink}{rgb}{1.0, 0.0, 0.5}
\hypersetup{ linktoc=all,
    colorlinks, linkcolor={palatinateblue},
    citecolor={brightpink}, urlcolor={amaranth}
}
\newcommand{\changeurlcolor}[1]{\hypersetup{urlcolor=#1}}
\graphicspath{{Images/}}

\begin{document}

\title{Entanglement Entropy and Complexity for One-Dimensional \\
Holographic Superconductors}
\author{Mahdi Kord Zangeneh}
\email{mkzangeneh@shirazu.ac.ir}
\affiliation{Center for Gravitation and Cosmology, College of Physical Science and Technology, Yangzhou University, Yangzhou 225009, China}
\affiliation{Research Institute for Astronomy and Astrophysics of Maragha
(RIAAM)-Maragha, IRAN, P. O. Box: 55134-441}
\affiliation{Physics Department and Biruni Observatory, Shiraz University, Shiraz 71454,
Iran}
\affiliation{Center for Astronomy and Astrophysics, Department of Physics and Astronomy,
Shanghai Jiao Tong University, Shanghai 200240, China}

\author{Yen Chin Ong}
\email{ongyenchin@gmail.com}
\affiliation{Center for Gravitation and Cosmology, College of Physical Science and Technology, Yangzhou University, Yangzhou 225009, China}
\affiliation{Center for Astronomy and Astrophysics, Department of Physics and Astronomy,
Shanghai Jiao Tong University, Shanghai 200240, China}
\author{Bin Wang}
\email{wang\_b@sjtu.edu.cn}
\affiliation{Center for Astronomy and Astrophysics, Department of Physics and Astronomy,
Shanghai Jiao Tong University, Shanghai 200240, China}
\affiliation{Center for Gravitation and Cosmology, College of Physical Science and Technology, Yangzhou University, Yangzhou 225009, China}

\begin{abstract}
Holographic superconductor is an important arena for holography, as it
allows concrete calculations to further understand the dictionary between
bulk physics and boundary physics. An important quantity of recent interest
is the holographic complexity. Conflicting claims had been made in the literature concerning the behavior of holographic complexity during phase transition. We clarify this issue by performing a numerical study on
one-dimensional holographic superconductor. Our investigation shows that 
holographic complexity does not behave in the same way as holographic entanglement entropy. 
Nevertheless, the universal terms of both quantities are finite and reflect the phase transition at the same critical temperature. \newline\newline
\end{abstract}

\maketitle

\section{Introduction: Holographic Complexity and Phase Transition}

AdS/CFT correspondence, or holography, has shown 
a deep connection between gravity in an asymptotically anti-de Sitter
(AdS) spacetime (``the bulk'') and the quantum field theory 
that lives on its conformal boundary \cite{maldacena}. In recent years,
quantum information has been applied in the context of gravitational
physics, notably in the context of black hole information paradox \cite{1409.1231}. 
Two quantities of the boundary field theory, which play important roles in quantum information, 
are entanglement entropy and (quantum) complexity.

As it turns out, both of these quantities are reflected in the bulk geometry. The entanglement entropy between the
degrees of freedom inside a closed region $\mathcal{A}$ with that of
its exterior (both of which are on the boundary), is proportional to the area of an extremal surface, 
$\gamma _{\mathcal{A}}$, that anchors on the boundary of $\mathcal{A}$, i.e., $\partial
\mathcal{A}=\partial \gamma _{\mathcal{A}}$, and extends into the bulk\footnote{This simple statement hides a lot of mathematical subtleties \cite{1312.3699}.} (if there are more than
one such surfaces, the one with minimal area is chosen). 
Specifically, the Ryu-Takayanagi formula \cite{ryu1,ryu2} states that the (regularized)
holographic entanglement entropy (HEE) is given by, in the units $c=\hbar
=k_{B}=1$, and $G$ being the Newton's constant,
\begin{equation}
\mathcal{S}=\frac{\text{Area}(\gamma _{\mathcal{A}})}{4G}.
\end{equation}%
The holographic complexity (HC) for this subregion $\mathcal{A}$ is conjectured to be holographically related to the
volume enclosed by the aforementioned minimal surface. Specifically \cite%
{alishahiha}, up to a constant factor (the factor $8\pi$ is merely a
convention), 
\begin{equation}
\mathcal{C}=\frac{\text{Volume}(\gamma _{\mathcal{A}})}{8\pi \mathcal{R}G},  \label{HC0}
\end{equation}%
where $\mathcal{R}$ is the radius of curvature of the background spacetime, e.g., the
AdS curvature radius, as in this work. Note that both $\mathcal{S}$ and $\mathcal{C}$ are dimensionless quantities in our choice of units.

Essentially, HEE is related to the content of information encoded in the subsystem (for example, information starts to leak out from a black hole during the Page time \cite{page1,page1b,page2}, which is the moment when the HEE of the Hawking radiation starts to decrease). On the other hand, HC has to do with how difficult it is to perform an operation. In the context of Hawking radiation, this is related to the difficulty of decoding and extracting the highly scrambled information from the Hawking radiation \cite{1301.4504,1301.4505,1403.4886}. In holography, HC of a field theory can be interpreted as the minimum number of gates to implement a certain unitary operator, to turn a pure reference state into another pure state \cite{1402.5674}. For mixed states the interpretation in terms of gates is not as straightforward, we will return to this in the next section.

Due to the ambiguity of choosing the correct length
scale $\mathcal{R}$ for different backgrounds, it has been recently proposed that one
should use instead the Einstein-Hilbert action in the so-called
\textquotedblleft Wheeler-DeWitt patch\textquotedblright\ as the holographic
dual of complexity. However, with the right choice of the length scale, the
original \textquotedblleft complexity=volume\textquotedblright\ conjecture
yields essentially the same result as that of the more recent
\textquotedblleft complexity=action\textquotedblright\ conjecture \cite%
{1509.07876,1512.04993}.

In this work we will focus on holographic complexity as defined in Eq. (\ref%
{HC0}) above, which remains the form that is widely focused on in the
holography literature. We will focus on the time-independent
subregion holographic complexity, i.e., the minimal surface $\gamma _{\mathcal{A}}$ is
entirely outside of the black hole horizon. (See \cite{1609.02514} for more discussions on the properties of such time-independent volume in various circumstances.)

Since complexity essentially measures the
difficulty of turning a quantum state into another, it is conceivable that a
phase transition on the boundary field theory could be reflected in the HC.
This possibility is further supported by the recent proposal that the
complexity is deeply connected with fidelity susceptibility \cite%
{1604.06909, 1702.06386, 1702.07471, 1507.07555}, which is known to be able to probe
phase transition, even without prior knowledge of the local order parameter 
\cite{1502.06969, 0701608, 0704.2945, 0811.3127}.

Our work was motivated by the discrepancies between existing results in the
literature: Momeni et al. \cite{momeni} claimed that during the
phase transition of a one dimensional holographic superconductor, there is a
divergent behavior in the HC. (This divergence should, of course, not to be confused with the trivial divergence that could be removed via regularization.) However, Roy and Sarkar found that as far as phase
transitions are concerned, HC captures essentially the same information as
HEE \cite{1701.05489}, which had previously been investigated in \cite{1306.4955}. 
 (For recent discussion regarding the relations
between HEE and HC, see, e.g., \cite{1703.01337}.)
Although the latter does not investigate a $1$-dimensional holographic superconductor, but rather a thermodynamics phase
transition of a Reissner-Nordstr\"{o}m-AdS black hole, it does suggest that the behavior of HC during phase transition should be carefully re-examined.

Furthermore, a divergence in the universal terms of HC during phase transition is rather problematic for the following reason.
Quantum complexity can be understood, in the language of circuits, as the minimum number of gates that is
required to implement a certain unitary operator, to turn a pure reference state into another pure state \cite{1402.5674}. 
For subregions, such as the one we discussed here, the states are mixed, and so its interpretation is somewhat subtle. Consider the density matrix $\rho_\mathcal{A}$ associated with the subregion $\mathcal{A}$. One then prepares $\rho_\mathcal{A}$ with a completely positive trace-preserving
(CPTP) map acting on the reference state \cite{1612.00433}. In doing so, a number of ``ancillary'' and ``erasure'' gates, which add and remove additional
degrees of freedom, are added to the circuits \cite{9806029,0804.3401}. Effectively, this means that we can interpret the subregion complexity in the following way:
one first extend the Hilbert space of $\mathcal{A}$ with new ancillary degrees of freedom, which would purify the mixed state $\rho_\mathcal{A}$. The subregion complexity is then the minimum number of (universal) gates required to turn a reference pure state into the required pure state \cite{1612.00433}.

This point of view would mean that during phase transition, the field theory becomes so complex
that one requires an infinite number of gates. So, if correct, the result of 
\cite{momeni} would mean that not only does HC respond to phase transition,
it also does so \emph{extremely drastically}. An infinite amount of complexity does not appear to be physically plausible. 

We therefore return to the model investigated in \cite{momeni}, which is a fully backreacted 1-dimensional holographic superconductor, and perform a numerical analysis to further investigate this issue. Indeed, it was mentioned in \cite{momeni} that such a numerical analysis is interesting and should be carried out to supplement their analytic analysis.

Our numerical investigation shows \emph{conclusively} that notwithstanding the claim of \cite{momeni}, during the phase transition of a 
$1$-dimensional holographic superconductor, the universal terms of HC remains finite and well-defined, just like HEE. 
In particular, both HC and HEE show that the superconducting phase intersect with the normal phase at the same critical temperature. 
We will further explain why our numerical result is inconsistent with \cite{momeni}.
However, in contrast to the claim made in \cite{1701.05489} that HC contains the same information as HEE as far as phase transitions are
concerned, we found that HC and HEE can behave quite differently. There is, however, no conflict with  the results in \cite{1701.05489} since the system studied therein is substantially different from ours.

\section{One-Dimensional Holographic Superconductor: The Set-Up\label{FE}}

In this section, we first introduce the background geometry of a black hole
coupled with a charged complex scalar field, and explain how to take
backreaction of the matter field into account to model a holographic
superconductor numerically. (Readers who are unfamiliar with holographic superconductors may consult \cite{horowitz} for details.)

The holographic setup of a one-dimensional\footnote{%
By one-dimensional, we meant one spatial dimension, i.e., the superconductor lives in
a (1+1)-dimensional spacetime.} superconductor involves an asymptotically
anti-de Sitter bulk geometry, which is governed by the ($2+1$)-dimensional
action \cite{ren,liu}%
\begin{eqnarray}
S=\int \text{d}^{3}x\sqrt{-g} &&\left[ \frac{1}{2\kappa ^{2}}\left( R+\frac{2%
}{l^{2}}\right) \right. -\frac{1}{4}F_{ab}F^{ab}  \notag \\
&&\left. -\left\vert \nabla \psi -iqA\psi \right\vert ^{2}-m^{2}\left\vert
\psi \right\vert ^{2}\right] .  \notag \\
&&
\end{eqnarray}%
Here $R$ and $g$ are, respectively, the Ricci scalar and the determinant of
the metric; $\kappa =\sqrt{8\pi G_{3}}$ is the $(2+1)$-dimensional
gravitational constant with $G_{3}$ being the $(2+1)$-dimensional Newton's
constant, and $l$ is the asymptotic AdS curvature radius. Also in the action
one finds the electromagnetic tensor $F_{ab}=\nabla_{\lbrack a}A_{b]}$,
where $A_{b}$ is the usual vector gauge potential. The gauge field is
coupled to a charged complex scalar field $\psi $, with $m$ and $q$ being
the mass and the charge of the scalar field, respectively.

In order to study the fully backreacted holographic superconductor, we
consider an ansatz of the form%
\begin{gather}\label{metric}
\text{d}s^{2}=-f\left( z\right) e^{-\chi \left( z\right) }\text{d}%
t^{2}+\left( z^{4}f\left( z\right) \right) ^{-1}\text{d}z^{2}+\left(
zl\right) ^{-2}\text{d}x^{2},
\end{gather}%
where $\left\{t,z,x\right\}$ are the usual Poincar\'e-type coordinates in
asymptotically AdS spacetime.

We are interested in static, translationally invariant solutions, thus we
consider the ansatz for the gauge potential and the charged scalar field to
be \cite{0803.3295}, respectively, 
\begin{equation}
{A}=\phi \left( z\right) \text{d}t,\text{ \ \ \ \ }\psi =\psi \left(
z\right) .
\end{equation}
Since the Maxwell equations imply that $\psi \left( z\right) $ has a
constant phase, it can be considered as a real function without loss of
generality. In this setup, the black hole horizon and the charge of scalar
field can be fixed as unity by virtue of scaling symmetries \cite%
{0810.1563,ren,liu}.

The dual one-dimensional superconductor lives on the boundary at $%
z=0$. It can be shown that, with backreaction governed by $\kappa$ which is
treated as a parameter\footnote{%
We would like to consider the backreaction of the bulk fields on the
background metric. The bulk fields are the scalar $\psi$ and the gauge
potential coefficient $\phi$. If one re-scales them to $q\psi$ and $q\phi$
in the action, then although the Maxwell and scalar equations remain
invariant, the gravitational coupling is re-scaled by $\kappa^2\mapsto
\kappa^2/q^2$. Fixing the charge, one can vary $\kappa$, which now serves
as a backreaction parameter. Note that the limit $\kappa \to 0$ corresponds
to the probe limit, i.e., there is no backreaction. Alternatively, if one
fixes $\kappa$, then $q\to \infty$ limit corresponds to the probe limit \cite%
{1009.1991,1205.3543}.}, the field equations read%
\begin{eqnarray}
0 &=&\psi ^{\prime \prime }+\psi ^{\prime }\left( \frac{f^{\prime }}{f}+%
\frac{1}{z}-\frac{\chi ^{\prime }}{2}\right) +\frac{\psi }{z^{4}f}\left( 
\frac{e^{\chi }\phi ^{2}}{f}-m^{2}\right) ,  \notag \\
&& \\
0 &=&\phi ^{\prime \prime }+\left( \frac{1}{z}+\frac{\chi ^{\prime }}{2}%
\right) \phi ^{\prime }-\frac{2\phi \psi ^{2}}{z^{4}f}, \\
0 &=&f^{\prime }-\kappa ^{2}\left[ \frac{2\psi ^{2}}{z^{3}}\left( m^{2}+%
\frac{\phi ^{2}e^{\chi }}{f}\right) \right.  \notag \\
&&\text{ \ \ \ \ \ \ \ \ \ \ }\left. +zf\left( 2\psi ^{\prime 2}+\frac{\phi
^{\prime 2}e^{\chi }}{f}\right) \right] +\frac{2}{l^{2}z^{3}}, \\
0 &=&\chi ^{\prime }-4\kappa ^{2}z\left( \frac{\phi ^{2}\psi ^{2}e^{\chi }}{%
z^{4}f^{2}}+\psi ^{\prime 2}\right) ,
\end{eqnarray}%
where prime denotes the derivative with respect to $z$.

Our aim in this letter is to study numerically the behaviors of holographic
entanglement entropy (HEE) and complexity (HC) for a subregion with length $%
L $ of a fully backreacted one-dimensional superconductor. We will consider
both the normal and superconducting phases. We will employ the shooting
method to carry out our numerical calculation. In order to do this, we need
to know the behaviors of the functions of the above setup at both the black hole
horizon and the boundary. At the horizon, we can Taylor expand the field
equations to find the expansion coefficients of the following functions%
\begin{eqnarray}
f\left( z\right) &=&f_{1}(1-z)+f_{2}(1-z)^{2}+\cdots ,  \notag \\
\psi \left( z\right) &=&\psi _{0}+\psi _{1}(1-z)+\psi _{2}(1-z)^{2}+\cdots ,
\notag \\
\phi \left( z\right) &=&\phi _{1}(1-z)+\phi _{2}(1-z)^{2}+\cdots ,  \notag \\
\chi \left( z\right) &=&\chi _{0}+\chi _{1}(1-z)+\chi _{2}(1-z)^{2}+\cdots .
\end{eqnarray}%
Note that in the expansions above, we impose the condition that the metric
function $f$ and the gauge potential $\phi $ should both vanish on the black
hole horizon. The latter is applied so that the norm of  the gauge potential, $%
A_{\mu }A^{\mu }$, is finite at the horizon.

In order to perform the shooting method, we will find the coefficients $\psi
_{0}$, $\phi _{1}$ and $\chi _{0}$ such that the desired values for some
parameters on the boundary is attained. In our study, we will focus on the
case $m=0$. For this case, the various functions at the boundary are
approximately given by%
\begin{gather}
\chi \approx \chi _{-},\text{ \ \ \ \ }f\approx \left( zl\right) ^{-2}, 
\notag \\
\phi \approx \rho +\mu \ln \left( z\right) ,\text{ \ \ \ \ }\psi \approx
\psi _{-}+\psi _{+}z^{2},
\end{gather}%
where $\chi _{-}$, $\rho $, $\mu $, $\psi _{-}$ and $\psi _{+}$ are some
constants. According to the holographic dictionary, $\mu $ corresponds to the chemical potential of the dual
superconductor. The quantity $\psi _{+}$ is related to the expectation value
of the order parameter $\left\langle \mathcal{O}_{+}\right\rangle $ of the
dual superconductor, whereas $\psi _{-}$ is considered as the source of this
order parameter.

\begin{figure*}[t]
\centering{%
\subfigure[~$\kappa^2=0.1$]{
   \label{fig1a}\includegraphics[width=.46\textwidth]{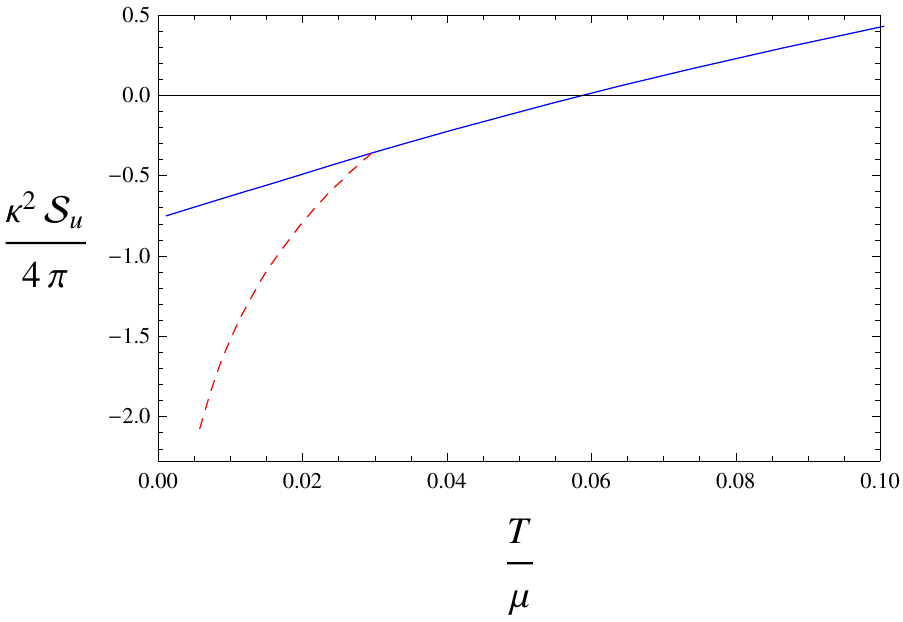}\qquad}}%
\subfigure[~$\kappa^2=0.2$]{
   \label{fig1b}\includegraphics[width=.46\textwidth]{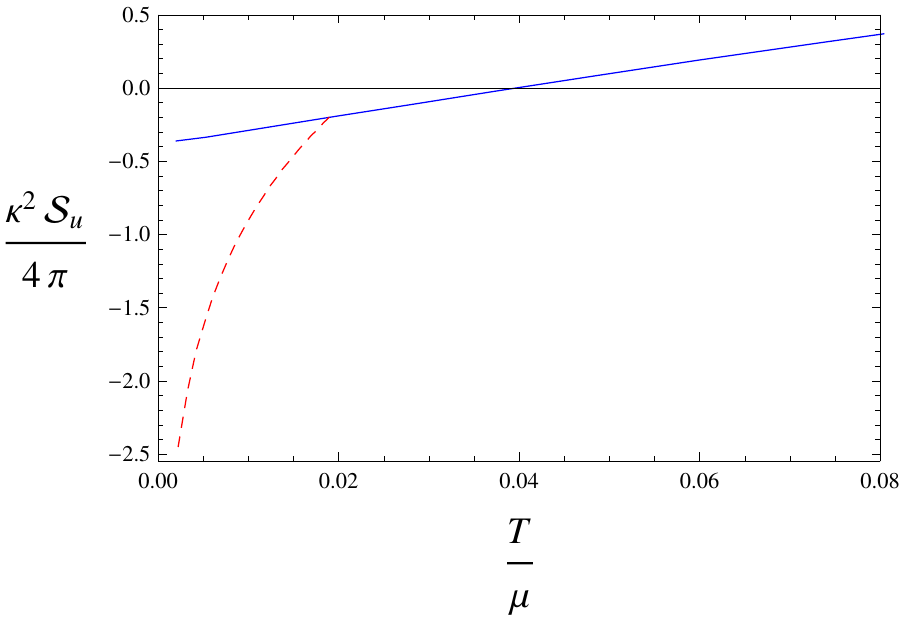}\qquad}
\caption{The behavior of $\protect\kappa ^{2}\mathcal{S}_{u}/4\protect\pi $
versus $T/\protect\mu $ for $l=1$ and $\protect\mu L/2=1$. Blue (solid) and
red (dashed) curves correspond respectively to normal and superconducting
phases. For $\protect\kappa ^{2}=0.1$ and $0.2$, the critical temperatures
per chemical potential are $T_{c}/\protect\mu =0.0295$ and $0.0189$,
respectively \protect\cite{liu}.}
\label{fig1}
\end{figure*}

\begin{figure*}[t]
\centering{%
\subfigure[~$\kappa^2=0.1$]{
   \label{fig2a}\includegraphics[width=.46\textwidth]{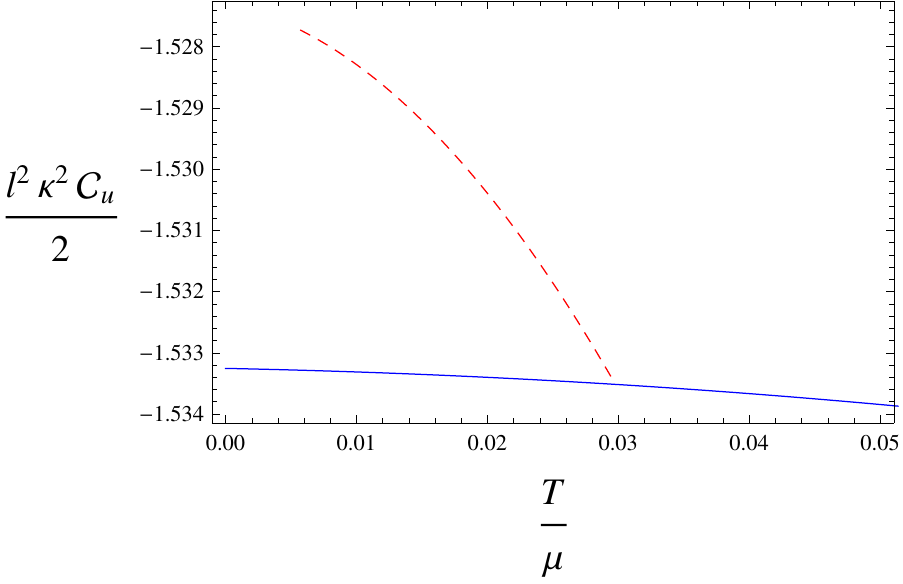}\qquad}}%
\subfigure[~$\kappa^2=0.2$]{
   \label{fig2b}\includegraphics[width=.46\textwidth]{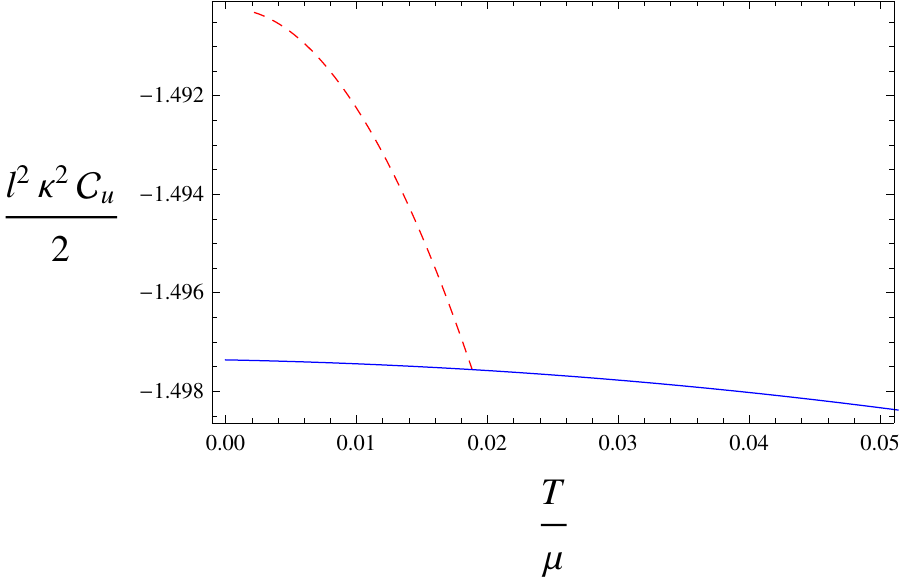}\qquad}
\caption{The behavior of $l^{2}\protect\kappa ^{2}\mathcal{C}_{u}/2$ versus $%
T/\protect\mu $ for $l=1$ and $\protect\mu L/2=1$. Blue (solid) and red
(dashed) curves correspond respectively to normal and superconducting phases.
For $\protect\kappa ^{2}=0.1$ and $0.2$, the critical temperatures per
chemical potential are $T_{c}/\protect\mu =0.0295$ and $0.0189$,
respectively \protect\cite{liu}.}
\label{fig2}
\end{figure*}

\begin{figure*}[t]
\centering{%
\subfigure[]{
   \label{fig2a}\includegraphics[width=.46\textwidth]{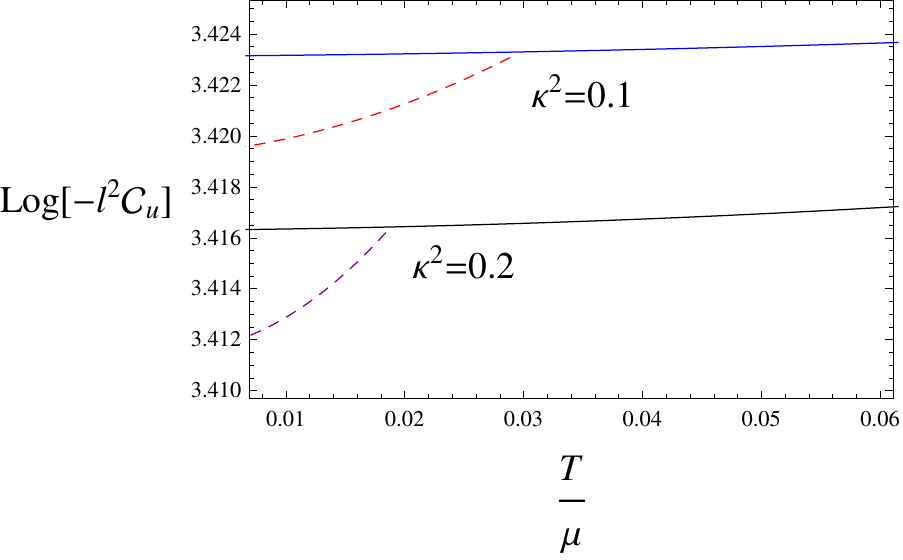}\qquad}}%
\subfigure[]{
   \label{fig2b}\includegraphics[width=.46\textwidth]{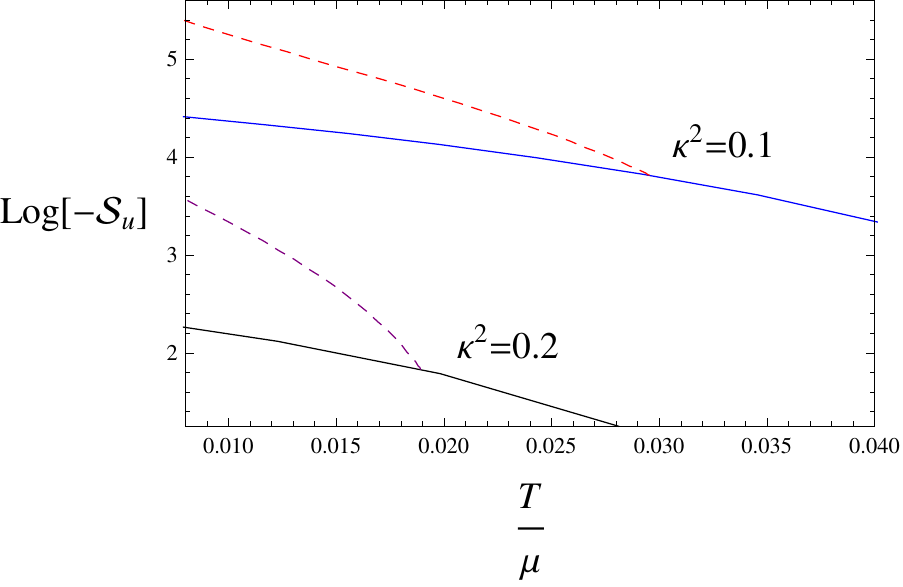}\qquad}
\caption{The logarithms of $-\l^2\mathcal{C}_u$ and $-\mathcal{S}_u$ plotted against $T/\mu$.
The top set of curves correspond to $\kappa^2=0.1$, while the lower set of curves  correspond to $\kappa^2=0.2$.
The curves for $\kappa^2=0.2$ have been shifted vertically to fit both set of curves into a single plot.}
\label{fig3}
\end{figure*}

To apply the shooting method, we change the value of the coefficient $\phi
_{1}$ and set $\psi _{0}$ and $\chi _{0}$ at the horizon so that both $\psi
_{-}$ and $\chi _{-}$ vanish at the boundary. The former is required since
we treat the superconducting phase transition as a spontaneous symmetry
breaking phenomenon (i.e., the symmetry breaking is entirely due to the low temperature, not induced by a source), while the latter is allowed by the rescaling symmetry%
\begin{equation}
e^{\chi }\rightarrow a^{2}e^{\chi },\text{ \ \ \ \ }\phi \rightarrow \phi /a,%
\text{ \ \ \ \ }t\rightarrow at,
\end{equation}
for some constant $a$.

The temperature of holographic superconductor is given by the Hawking
temperature of the black hole in the bulk \cite{momeni, liu}, which can be easily checked by the usual method of Wick-rotating the metric in Eq. (\ref{metric}) to Euclidean signature and imposing regularity at the Euclidean horizon:
\begin{equation}
T=\left. \frac{e^{-\chi /2}f^{\prime }}{4\pi }\right\vert _{z=1}.
\end{equation}%
The HEE $\mathcal{S}$ associated to the subregion $\mathcal{A}$ of a dual field theory
is proportional to the area of a minimal co-dimension two surface $\gamma
_{\mathcal{A}}$ such that their boundaries are the same, i.e., $\partial \gamma
_{\mathcal{A}}=\partial \mathcal{A}$ \cite{ryu1,ryu2}. Therefore, for a strip subregion with
length $L$, we have \cite{momeni}%
\begin{equation}
\mathcal{S}=\frac{2\pi }{\kappa ^{2}}\int_{-L/2}^{L/2}\frac{\text{d}x}{z^{2}}%
\sqrt{\frac{1}{f}\left( \frac{\text{d}z}{\text{d}x}\right) ^{2}+\frac{z^{2}}{%
l^{2}}}.  \label{HEE}
\end{equation}%
The minimality condition implies%
\begin{equation}
\frac{\text{d}z}{\text{d}x}=\pm l^{-1}\sqrt{\left( z_{\ast
}^{2}-z^{2}\right) f},  \label{dzdx}
\end{equation}%
in which the constant $z_{\ast }$ satisfies the stationary condition $\left. 
\text{d}z/\text{d}x\right\vert _{z=z_{\ast }}=0$. This can be verified using
the Euler-Lagrange variation method. Next, setting $x_{\mp }\left( z_{\ast
}\right) =0$, we find 
\begin{equation}
x_{\mp }\left( z\right) =\mp l\int_{z}^{z_{\ast }}\frac{\text{d}z}{\sqrt{%
\left( z_{\ast }^{2}-z^{2}\right) f}}.  \label{xz}
\end{equation}%
This satisfies, with a UV cutoff $\epsilon$,
\begin{equation}
x_{\mp }\left( \epsilon \rightarrow 0\right) =\mp L/2. \label{z*}
\end{equation}%
Notice that $+$ (respectively, $-$) in Eq. (\ref{dzdx}) corresponds to the region $%
-L/2<x\leqslant 0$ (respectively, $0\leqslant x<L/2$) while in Eq. (\ref{xz}), it
corresponds to $0\leqslant x<L/2$ (respectively, $-L/2<x\leqslant 0$). Using Eq. (\ref%
{dzdx}), one can rewrite Eq. (\ref{HEE}) as%
\begin{equation}
\mathcal{S}=\frac{4\pi }{\kappa ^{2}}\int_{\epsilon \rightarrow 0}^{z_{\ast
}}\frac{z_{\ast }\text{d}z}{z^{2}l\sqrt{\left( z_{\ast }^{2}-z^{2}\right) f}}%
.  \label{HEE2}
\end{equation}%
On the other hand, the complexity corresponding to $\mathcal{A}$ is holographically related to the volume
in the bulk enclosed by $\gamma _{\mathcal{A}}$, namely \cite{momeni,alishahiha}%
\begin{equation}
\mathcal{C}=\frac{2}{l^{2}\kappa ^{2}}\int_{\epsilon \rightarrow 0}^{z_{\ast
}}\frac{x_{+}\left( z\right) \text{d}z}{z^{3}\sqrt{f}}.  \label{HC}
\end{equation}%
Notice that, according to the scaling symmetry 
\begin{equation}
\left( t,z,x\right) \rightarrow b^{-1}\left( t,z,x\right) ,\text{ \ \ \ \ }%
f\rightarrow b^{2}f,\text{ \ \ \ \ }\phi \rightarrow b\phi ,
\end{equation}%
the quantities $T$, $L$, $\mu $, $\mathcal{S}$ and $\mathcal{C}$, scale as%
\begin{gather}
T\rightarrow bT,\text{ \ \ \ \ }L\rightarrow b^{-1}L,\text{ \ \ \ \ }\mu
\rightarrow b\mu ,  \notag \\
\mathcal{S}\rightarrow \mathcal{S},\text{ \ \ \ \ }\mathcal{C}\rightarrow 
\mathcal{C}.
\end{gather}%
Therefore, to study the physics, it is useful to employ the dimensionless
quantities $T/\mu $, $\mu L$, $\mathcal{S}$ and $\mathcal{C}$.

At this point, let us give some comments about the formally diverging terms of HEE
and HC before regularization. The diverging term of HEE (Eq. (\ref{HEE2})) caused by the pure AdS
geometry $f\rightarrow \left( zl\right) ^{-2}$ near the UV cutoff $\epsilon $
is $\left( 4\pi \ln \epsilon ^{-1}\right) /\kappa ^{2}$. Subtracting this
diverging term from $\mathcal{S}$ in Eq. (\ref{HEE2}), one can find the
universal term of HEE, $\mathcal{S}_{u}$. As for the HC, the diverging term
corresponding to the pure AdS geometry is $2z_{\ast }/\left( l^{2}\kappa
^{2}\epsilon \right) $; it \emph{cannot} be used to subtract off the divergence in every
situation. For instance, for normal phase ($\psi =\chi =0$), the diverging
term includes $\tanh ^{-1}\left( z_{\ast }\right) /\epsilon $ for the $\kappa
\rightarrow 0$ case (see Appendix \ref{app1}). Indeed, the diverging term may include different \emph{%
functions} of $z_{\ast }$ under different situations. It is not possible to find a general form for HC
diverging term analytically. Fortunately, this is not necessary, since we
can overcome this problem numerically. The HC includes a universal term $%
\mathcal{C}_{u}$ and a diverging term in the form of $\mathcal{F}\left(
z_{\ast }\right) /\epsilon $. Since the value of universal term should not
change for different cutoffs, subtracting HC in Eq. (\ref{HC}) for two
different values of cutoff $\epsilon _{1}$ and $\epsilon _{2}$, one finds $%
\left( \epsilon _{1}^{-1}-\epsilon _{2}^{-1}\right) \mathcal{F}\left(
z_{\ast }\right) $. Therefore, the value of $\mathcal{F}\left( z_{\ast
}\right) $ in different situations can be found numerically. (The regularization of HC is discussed in great details in \cite{1701.03706}. See also \cite{1612.00433} for a discussion on the geometry related to the UV divergence in the HC.)

In the rest of this letter, we will study the behavior of (the universal parts of) HEE and HC of a 1-dimensional superconductor numerically. To do
this, we will first evaluate $z_{\ast }$ numerically using Eq. (\ref{z*}).
Then, we will obtain $x_{+}\left( z\right) $ numerically from Eq. (\ref{xz})
and from this, calculate HC from Eq. (\ref{HC}). We can also compute the HEE
given by Eq. (\ref{HEE2}).

\section{Numerical results for HEE and HC\label{num}}

In this section, we will study HEE and HC for a strip subregion $\mathcal{A}$ of the
1-dimensional dual system. We first show below the numerical results: the
plots of the universal terms of HEE, $\mathcal{S}_{u}$ (Fig. \ref{fig1}),
and HC, $\mathcal{C}_{u}$ (Fig. \ref{fig2}), against the ratio of temperature to chemical potential, $T/\mu$,  for
both normal and superconducting phases.

As mentioned in the Introduction, we are motivated by the inconsistency in the literature:
on one hand is the claim by Momeni et al. \cite{momeni} that during the
phase transition of a $1$-dimensional holographic superconductor, there is a
divergent behavior in HC.  On the other hand, Roy and Sarkar found that during the phase transition of a Reissner-Nordstr\"{o}m-AdS black hole, 
HC behaves in the same manner as HEE \cite{1701.05489}. Granted that the latter does not investigate a 1-dimensional holographic superconductor, 
it is quite suggestive that there is a conflict between the two results. 

Here, our numerical investigation shows \emph{conclusively} that, during phase transition of a 
$1$-dimensional holographic superconductor, the universal terms of holographic complexity is still finite and well-defined. 
Indeed, as can be seen from Fig. \ref{fig1}, the points where the
plots of $\mathcal{S}_{u}$ for the normal phase (blue solid curves)
intersect with that of the superconducting phase (red dashed curves), occur
at critical temperatures $T_{c}/\mu =0.0295$ and $T_{c}/\mu =0.0189$, for backreaction parameters $%
\kappa ^{2}=0.1$ and $\kappa ^{2}=0.2$, respectively. This agrees with the
results in Fig. \ref{fig2}, which are plots for the universal terms in HC. Varying $\kappa^2$ does not change the qualitative behavior of these plots.

In other words, the critical temperature of the phase transition can be read from the plot of $\mathcal{C}_u$,  
which agrees with the critical temperature read from the plot of $\mathcal{S}_u$. 
This means that HC does indeed responds to phase transitions, just like the HEE would.
It is worth noting that increasing the strength of backreaction makes condensation harder, i.e., it occurs at a lower temperature. 

In Fig. \ref{fig3}, we have also plotted the logarithms of $-\l^2\mathcal{C}_u$ and $-\mathcal{S}_u$ as functions of $T/\mu$. The aim here is to show that the opening angles between the normal phase and superconducting phase (for both HEE and HC) increases as we increase the strength of backreaction, $\kappa^2$. (In order to fit both set of curves into a single plot, we have shifted the curves for $\kappa^2=0.2$ vertically.)  Thus, for a fixed value of $T/\mu$, a larger value of $\kappa^2$ means that there is a larger difference between the values of HEE and HC of the normal phase compared to those of the superconducting phase.

We now compare our results to Roy and Sarkar \cite{1701.05489},  and Momeni et al. \cite{momeni}.
Roy and Sarkar found that HC contains the same information as HEE, as far as phase transitions are concerned. To be more precise, they 
investigated the phase transition of a (3+1)-dimensional spherical Reissner-Nostr\"om AdS black hole in Section 5 of their work \cite{1701.05489}, in which they
plotted the graphs of renormalized complexity for fixed charge ensemble and fixed opening angle $\theta_0$ (the entangling region being a spherical cap defined by  $\theta \leqslant \theta_0$). It turned out that complexity behaves in the same way as entanglement entropy, whose plots are shown in Section 5 of \cite{1306.4955}. In our case however, it is clear that Fig. \ref{fig1} are not similar to Fig. \ref{fig2}: $\mathcal{S}_u$ increases with $T/\mu$, while $\mathcal{C}_u$ decreases with $T/\mu$. So our result shows that HC and HEE need \emph{not} behave in the same manner in the context of phase transitions (in either the normal phase or the  superconducting phase). This does not contradict the results in \cite{1701.05489} since there are a few differences between our set-up and theirs: our bulk spacetime is (2+1)-dimensional, and our subsystem is strip-shaped, whereas in  \cite{1701.05489} they considered a (3+1)-dimensional bulk with spherical horizon, and furthermore their subsystem is circular. Note that HC and HEE behave differently in our work \emph{even during normal phase}, so this suggests that the differences between our results and that of \cite{1701.05489} is not simply due to us considering a different kind of phase transition (superconductor instead of a thermodynamical one).

On the other hand, our result, which shows no divergence in the behavior of HC, is clearly inconsistent with Momeni et al. \cite{momeni}.
Going back to their analytic calculation, we found that their analysis was not performed carefully. The
expression of HC is, up to a positive dimensionful factor, of the form (see Eq. ($28$) of their work for the full expression)
\begin{equation}
\mathcal{C}|_{T\to T_c} \sim  \left[\frac{1}{\mu - \mu_c} \left(\frac{T}{T_0}-1\right)+\text{const.}\right]^2 + \cdots,  \label{memoni1}
\end{equation}%
which they claimed to be divergent in the limit $\mu \rightarrow \mu _{c}$.
Here $T_{0}$ is the Hawking temperature of a pure BTZ black hole. However,
according to the discussion below Eq.(24) of their paper, one also has $T_{0}\rightarrow
T_{c}$ near the critical point, so that in the phase transition
limit the expression above becomes indeterminate. One
therefore cannot conclude whether there is a divergent behavior from this
analysis alone, though it potentially could still happen. 

Our numerical work shows
that this does not happen. Indeed, the location of the black hole horizon is, up to a positive dimensionful factor\footnote{In our setup, as mentioned in the previous section, we have fixed both the horizon and charge of the scalar field to be unity by utilizing scaling symmetries. Thus, Eq. (\ref{memoni1}) only applies to the work of \cite{momeni}, not the ones carried out in this paper.}, 
\begin{equation}
z_+ \sim  \left[\frac{1}{\mu - \mu_c} \left(\frac{T}{T_0}-1\right)+\text{const.}\right]^{-1}.  \label{memoni1}
\end{equation}%
(See Eq. ($27$) of \cite{momeni}; but with their coordinate $r=1/z$). This expression will tend to zero if HC is indeed
divergent. However, to be well-defined in the context of holography, the horizon should be well inside the bulk and
thus $z_+$ should always be bounded away from 0. This is only possible if the potential divergence is avoided due to an
indeterminate form as remarked above. 

\section{Summary and discussion}

In this work, we conducted a numerical analysis of the holographic
complexity (HC) and the holographic entanglement entropy (HEE) for a
fully-backreacted $1$-dimensional holographic superconductor. We showed that
both quantities reflect the presence of a phase transition. We found no
divergent behavior in the HC during phase transition, contrary to
the claim in \cite{momeni}, whose analytic analysis contains a mistake. Despite these mistakes, the analytic work of \cite{momeni} is nontrivial and important.

Furthermore, our results demonstrated that in the context of a 1-dimensional holographic superconductor, the universal part of the entanglement entropy, $\mathcal{S}_u$, is increasing with $T/\mu$. On the other hand, the universal part of the complexity, $\mathcal{C}_u$, is a decreasing function of $T/\mu$. Therefore, these two quantities behave quite differently for phase transitions that involve a superconducting phase, in contrast to thermodynamical phase transitions that were investigated in \cite{1701.05489}, in which HC and HEE behave, qualitatively, in the same manner. 

Some physical interpretations are useful at this point. During the normal phase, as temperature decreases, order in the system increases due to cooling, and so the effective degrees of freedom also decreases. This is consistent with the decrease in the entanglement entropy, since entanglement entropy \emph{is} a measure of the degrees of freedom in the field theory. Note that in the normal phase, the order parameter $\langle \mathcal{O}_+ \rangle$ is zero because $\psi$ is zero. The quantity $\langle \mathcal{O}_+ \rangle$ only determines the condensation in the superconducting phase.
For superconducting phase, HEE decreases as $T/\mu$ decreases because HEE is related to the degrees of freedom in the field theory, and so as Cooper pairs formed, it is expected that HEE would decrease. For the same reason, the HEE for the superconducting phase is lower than that of the normal phase \cite{1401.5720}. 

On the other hand, HC is related to the number of unitary operators that are required to reach some quantum state \cite{1402.5674}. A larger (but finite) complexity at lower temperatures is therefore related to the quantum state of the system becoming more complicated towards the critical temperature. The detailed underlying physics remains to be investigated. Regardless, both $\mathcal{C}_u$ and $\mathcal{S}_u$ reflect the presence of superconducting phase transition, \emph{at the same critical temperature}. We also found that, with $T/\mu$ fixed, increasing $\kappa^2$ means that there is a larger difference between the values of HEE and HC of the normal phase compared to the superconducting phase. See Fig. \ref{fig3}.

These findings indicate that HC and HEE can behave in different ways,  in both the normal phase as well as the superconducting phase. Since the results of \cite{1701.05489} showed that during a thermodynamical phase transition of a different system, HC and HEE \emph{do} behave in the same manner, it would be interesting to further investigate the sufficient and necessary conditions for HC and HEE to behave in the same manner, as well as the effects of different spacetime dimensions and the geometry of the underlying subsystem on the behaviors of HC and HEE. 

Lastly, it would also be interesting to compare the behavior of HC to that of fidelity susceptibility during phase transitions of various systems, so as to further investigate the  recently proposed connection between holographic complexity and (reduced) fidelity susceptibility \cite{1702.06386,1702.07471}, or ``RFS/HC duality'' for short.

Holographic entanglement entropy has proven to be a useful concept, which has allowed us to further understand the quantum information theoretic aspects of gravity and holography. We expect that holographic complexity, being a relatively new and novel concept, has the potential to offer even more nontrivial insights into the deep and subtle connection between gravity and field theory \cite{1411.0690}. 

\begin{acknowledgments}
MKZ would like to thank Shanghai Jiao Tong University for the warm
hospitality during his visit. He also thanks the Research Council of Shiraz University.
The work of MKZ has been supported financially by Research Institute for Astronomy \& Astrophysics of Maragha (RIAAM) under research project No. 1/4717-169.
YCO and BW acknowledge the support by the National Natural
Science Foundation of China (NNSFC).
The authors thank Xiao-Mei Kuang for useful discussions.
\end{acknowledgments} 

\appendix

\section{HC for normal phase without backreaction\label{app1}}

In this appendix, we shall discuss the diverging term of the holographic complexity for a strip-shaped subregion.
Related discussions concerning a ball-shaped subregion can be found in, e.g., \cite{alishahiha,1702.07471}.

Setting $l=1$, let us consider the normal phase case ($\psi =\chi =0$), of which we know
the solution,%
\begin{equation}
f\left( z\right) =z^{-2}-1+\kappa ^{2}\mu ^{2}\ln \left( z\right) \text{ \
and \ }\phi \left( z\right) =\mu \ln \left( z\right) .  \notag
\end{equation}%
In the probe limit $\kappa \rightarrow 0$, i.e., in the absence of backreaction,
one finds:

\begin{eqnarray}
x_{+}\left( z\right)  &=&\int_{z}^{z_{\ast }}\frac{\text{d}z}{\sqrt{\left(
z_{\ast }^{2}-z^{2}\right) f}}=\int_{z}^{z_{\ast }}\frac{\text{d}z}{\sqrt{%
\left( z_{\ast }^{2}-z^{2}\right) \left( z^{-2}-1\right) }}  \notag \\
&=&\coth ^{-1}\left( \sqrt{\frac{1-z^{2}}{z_{\ast }^{2}-z^{2}}}\right) .
\label{HC1}
\end{eqnarray}%
Notice that $x_{+}\left( z_{\ast }\right) =0$ as required. By setting the
strip length as $L$, we can find $z_{\ast }$ via the boundary condition $x_{+}\left(
\epsilon \rightarrow 0\right) =L/2$. Finally, using Eq. (\ref{HC1}), we can
calculate HC as%
\begin{eqnarray}
\frac{l^{2}\kappa ^{2}\mathcal{C}}{2} &=&\int_{\epsilon \rightarrow
0}^{z_{\ast }}\frac{x_{+}\left( z\right) \text{d}z}{z^{3}\sqrt{f}}%
=\int_{\epsilon \rightarrow 0}^{z_{\ast }}\frac{\coth ^{-1}\left( \sqrt{%
\frac{1-z^{2}}{z_{\ast }^{2}-z^{2}}}\right) \text{d}z}{z^{3}\sqrt{\left(
z^{-2}-1\right) }}  \notag \\
&=&-\frac{\pi }{2}+\frac{\tanh ^{-1}\left( z_{\ast }\right) }{\epsilon }%
+O\left( \epsilon \right) .
\end{eqnarray}%
If one performs the same procedure for pure AdS background, in which $f=z^{-2}$, one will
find the diverging term in this case to be $z_{\ast }/\epsilon $, which is
obviously different from the result above. 

This shows that the diverging term of HC for a generic asymptotically AdS geometry is not the same as
that of a pure AdS spacetime. In the case of nonvanishing $\kappa $, we have
a logarithmic term in $f\left( z\right) $, and so it is not possible to find
the diverging term analytically. The same problem also arises in the
superconducting phase. In Sec. \ref{FE}, we have explained how to overcome
this problem numerically.

\end{document}